\documentclass[twocolumn,showpacs,preprintnumbers,amsmath,amssymb,superscriptaddress,nofootinbib,english]{revtex4-1}
\usepackage{times,amsmath,amsfonts,amssymb,epstopdf}
\usepackage{graphicx}
\usepackage{dcolumn}
\usepackage{bm}
\usepackage{float}
\usepackage{epsfig}
\usepackage{graphicx}
\usepackage{hyperref}
\usepackage[usenames]{color}
\usepackage{url}
\usepackage[normalem]{ulem}
\usepackage[T1]{fontenc}

\usepackage[usenames]{color}

\def\be{\begin{equation}}
\def\ee{\end{equation}}
\def\ben{\begin{eqnarray}}
\def\een{\end{eqnarray}}
\def\ba{\begin{array}}
\def\ea{\end{array}}

\newcommand{\bq}{\begin{eqnarray}}
\newcommand{\eq}{\end{eqnarray}}
\newcommand{\bes}{\begin{subequations}}
\newcommand{\ees}{\end{subequations}}

\begin{document}
\newcommand{\half}{{\textstyle\frac{1}{2}}}
\allowdisplaybreaks[3]
\def\triangledown{\nabla}
\def\grad3{\hat{\nabla}}
\def\a{\alpha}
\def\b{\beta}
\def\g{\gamma}\def\G{\Gamma}
\def\d{\delta}\def\D{\Delta}
\def\ep{\epsilon}
\def\et{\eta}
\def\z{\zeta}
\def\t{\theta}\def\T{\Theta}
\def\l{\lambda}\def\L{\Lambda}
\def\m{\mu}
\def\f{\phi}\def\F{\Phi}
\def\n{\nu}
\def\r{\rho}
\def\s{\sigma}\def\S{\Sigma}
\def\ta{\tau}
\def\x{\chi}
\def\o{\omega}\def\O{\Omega}
\def\k{\kappa}
\def\pa {\partial}
\def\ov{\over}
\def\br{\\}
\def\ud{\underline}

\def\nucubic{{\nu} \rm{Galileon}}
\def\nulcdm{{\nu} \Lambda\rm{CDM}}

\newcommand\lsim{\mathrel{\rlap{\lower4pt\hbox{\hskip1pt$\sim$}}
    \raise1pt\hbox{$<$}}}
\newcommand\gsim{\mathrel{\rlap{\lower4pt\hbox{\hskip1pt$\sim$}}
    \raise1pt\hbox{$>$}}}
\newcommand\esim{\mathrel{\rlap{\raise2pt\hbox{\hskip0pt$\sim$}}
    \lower1pt\hbox{$-$}}}
\newcommand{\dpar}[2]{\frac{\partial #1}{\partial #2}}
\newcommand{\sdp}[2]{\frac{\partial ^2 #1}{\partial #2 ^2}}
\newcommand{\dtot}[2]{\frac{d #1}{d #2}}
\newcommand{\sdt}[2]{\frac{d ^2 #1}{d #2 ^2}}

\title{$\nu$Galileon: modified gravity with massive neutrinos as a testable alternative to $\Lambda$CDM}

\author{Alexandre Barreira}
\affiliation{Institute for Computational Cosmology, Department of Physics, Durham University, Durham DH1 3LE, U.K.}
\affiliation{Institute for Particle Physics Phenomenology, Department of Physics, Durham University, Durham DH1 3LE, U.K.}

\author{Baojiu Li}
\affiliation{Institute for Computational Cosmology, Department of Physics, Durham University, Durham DH1 3LE, U.K.}

\author{Carlton M. Baugh}
\affiliation{Institute for Computational Cosmology, Department of Physics, Durham University, Durham DH1 3LE, U.K.}

\author{Silvia Pascoli}
\affiliation{Institute for Particle Physics Phenomenology, Department of Physics, Durham University, Durham DH1 3LE, U.K.}

\preprint{IPPP/14/ 27 DCPT/14/ 54}

\begin{abstract}

We show that, in the presence of massive neutrinos, the Galileon gravity model provides a very good fit to the current CMB temperature, CMB lensing and BAO data. This model, which we dub $\nucubic$, when assuming its stable attractor background solution, contains the same set of free parameters as $\Lambda\rm{CDM}$, although it leads to different expansion dynamics and nontrivial gravitational interactions. The data provide compelling evidence ($\gtrsim 6\sigma$) for nonzero neutrino masses, with $\Sigma m_\nu \gtrsim 0.4\ {\rm eV}$ at the $2\sigma$ level. Upcoming precision terrestrial measurements of the absolute neutrino mass scale therefore have the potential to test this model. We show that CMB lensing measurements at multipoles $l \lesssim 40$ will be able to discriminate between the $\nucubic$ and $\Lambda\rm{CDM}$ models. Unlike $\Lambda\rm{CDM}$, the $\nu$Galileon model is consistent with local determinations of the Hubble parameter. The presence of massive neutrinos lowers the value of $\sigma_8$ substantially, despite of the enhanced gravitational strength on large scales.  Unlike $\Lambda\rm{CDM}$, the $\nucubic$ model predicts a negative ISW effect, 
which is difficult to reconcile with current observational limits.

\end{abstract} 
\maketitle

\section{Introduction}

The Galileon gravity model, proposed by Refs.~\cite{PhysRevD.79.064036, PhysRevD.79.084003, Deffayet:2009mn}, offers an alternative to the concordance $\Lambda\rm{CDM}$ model to explain the late time cosmic acceleration. In this model, a scalar field (dubbed the Galileon) drives sizeable modifications to gravity on large scales, which can nevertheless be suppressed near massive bodies by the Vainshtein mechanism \cite{Vainshtein1972393}. The latter allows the theory to pass the stringent Solar System tests of gravity \cite{Will:2014kxa}. The so-called Quartic and Quintic sectors of the Galileon model suffer from a number of theoretical and observational complications. These include the relatively small energy cutoff below which the theory is phenomenologically well defined \cite{PhysRevD.79.064036, Burrage:2012ja}, and possible time variations of the effective gravitational strength on Solar System scales that cannot be suppressed by the Vainshtein mechanism \cite{Babichev:2011iz, Kimura:2011dc, Barreira:2013xea, Li:2013tda}. Here, we focus on the portion of the parameter space that avoids these problems, which is known as the Cubic Galileon model. This model was, however, thought to be unable to fit the currently available observational data, due to its strong Integrated Sachs-Wolfe (ISW) effect, enhanced matter clustering on large scales \cite{Barreira:2013eea, Kimura:2011td} and difficulties in matching the position of the Baryonic Acoustic Oscillation (BAO) peak.

The above-mentioned observational tensions are seen in Galileon models where neutrinos are treated as massless particles. However, the inclusion of massive neutrinos in cosmological studies should be mandatory following the detection of neutrino flavour oscillations in solar, atmospheric and reactor experiments \cite{GonzalezGarcia:2012sz}. These have placed bounds on the mass-squared differences of the three neutrino species, which imply $\Sigma m_\nu > 0.06\ {\rm eV}$ for a normal mass ordering, and $\Sigma m_\nu > 0.1\ {\rm eV}$ for an inverted mass ordering ($\Sigma m_\nu$ is the sum of the three neutrino masses). Currently, the most stringent upper bounds on $\Sigma m_\nu$ come from cosmological observations, although these are highly model and dataset dependent. Here, we investigate the impact that massive neutrinos have on Galileon gravity cosmologies. Our main conclusion is that {\it the presence of sufficiently massive neutrinos in Galileon gravity models results in an alternative cosmological scenario to $\Lambda${\rm CDM} that is consistent with the currently available cosmological data, and that is testable by future cosmological and laboratory experiments.}

\section{Background} 

The Einstein-Hilbert action of the Cubic Galileon model is given by
\bq\label{Galileon action}
&& S = \int {\rm d}^4x\sqrt{-g} \left[ \frac{R}{16\pi G} - \frac{1}{2}c_2\mathcal{L}_2 - \frac{1}{2}c_3\mathcal{L}_3 - \mathcal{L}_m\right],
\eq
where $g$ is the determinant of the metric $g_{\mu\nu}$, $R$ is the Ricci scalar, the model parameters $c_2$ and $c_3$ are dimensionless constants, and $\mathcal{L}_2$ and $\mathcal{L}_3$ are given by
\bq\label{L's}
\mathcal{L}_2 = \nabla_\mu\varphi\nabla^\mu\varphi,\ \ \ \ \ \ \ \  \mathcal{L}_3 = \frac{2}{M^3}\Box\varphi\nabla_\mu\varphi\nabla^\mu\varphi,
\eq
in which $\varphi$ is the Galileon field and $M^3\equiv M_{\rm Pl}H_0^2$, where $M_{\rm Pl}$ is the reduced Planck mass and $H_0 = 100h\ {\rm km/s/Mpc}$ is the present-day Hubble expansion rate (the subscript "$_0$" denotes present-day values). The action of Eq.~(\ref{Galileon action}) is invariant under the Galilean shift transformation $\partial_\mu\varphi \rightarrow \partial_\mu\varphi + b_\mu$ (for constant $b_\mu$), hence the name of the model. The interested reader can find the Einstein and Galileon field equations as Eqs.~(4), (5), (6) and (7) of Ref.~\cite{Barreira:2013eea}. The Friedmann equation is given by $3H^2 = \kappa\left[\bar{\rho}_r + \bar{\rho}_m + \bar{\rho}_\nu+ \bar{\rho}_{\varphi}\right]$, where $\kappa = M_{\rm Pl}^{-2} = 8\pi G$ and $\bar{\rho}_{r}$, $\bar{\rho}_{m}$, $\bar{\rho}_{\nu}$ are, respectively, the background energy density of radiation, matter (baryons and cold dark matter) and massive neutrinos; the background energy density and pressure of the Galileon field are given, respectively, by
\bq
\label{eq:density-background} \kappa\bar{\rho}_\varphi = \frac{c_2\dot{\varphi}^2}{2} + \frac{6c_3\dot{\varphi}^3H}{H_0^2};\ \ \  \kappa\bar{p}_\varphi = \frac{c_2\dot{\varphi}^2}{2} - \frac{2c_3\ddot{\varphi}\dot{\varphi}^2}{H_0^2},
\eq
and $w_{\varphi} = \bar{p}_\varphi/\bar{\rho}_\varphi$ is the Galileon field equation-of-state parameter (an overdot denotes a partial derivative w.r.t. physical time). The background Galileon field equation of motion is given by
\bq\label{eq:background-EoM}
0 &=& c_2\left[\ddot{\varphi} +3\dot{\varphi}H \right] + \frac{6c_3}{H_0^2} \left[ 2\ddot{\varphi}\dot{\varphi}H + 3\dot{\varphi}^2H^2 + \dot{\varphi}^2\dot{H}\right].
\eq
Note that in writing Eqs.~(\ref{eq:density-background}) and (\ref{eq:background-EoM}) we have made the substitution $\varphi/M_{\rm{Pl}} \rightarrow \varphi$. Reference \cite{DeFelice:2010pv} showed that different initial conditions of the Galileon background equations eventually merge into a common time evolution called a {\it tracker} solution. The latter is characterized by the relation $\dot{\varphi}H = {\rm constant} \equiv \xi H_0^2$, where $\xi$ is a dimensionless constant.  In Refs.~\cite{Barreira:2012kk} and \cite{Barreira:2013jma}, it was found that, in order for the Galileon models to fit the low-$l$ Cosmic Microwave Background (CMB) data, the background evolution should reach the tracker well before the onset of the accelerated expansion. Before this epoch, the impact of the Galileon field is negligible. As a result and without any loss of generality, when constraining the Galileon model, we can assume that the background follows the tracker at all cosmological epochs. Assuming a spatially flat universe (i.e., a vanishing curvature density, $\Omega_k = 0$), the expansion rate on the tracker is given analytically by
\bq\label{eq:tracker_H}
&&\left(H(a)/H_0\right)^2 = \frac{1}{2}\left[\Omega_{m0}a^{-3} + \Omega_{r0}a^{-4} + \Omega_{\nu}(a)\right] \nonumber \\
&& + \frac{1}{2}\sqrt{\left[\Omega_{m0}a^{-3} + \Omega_{r0}a^{-4} + \Omega_{\nu}(a)\right]^2 + 4\Omega_{\varphi0}},
\eq
where $\Omega_\nu(a) = \Omega_{\nu 0}\bar{\rho}_{\nu}(a)/\bar{\rho}_{\nu 0}$, $\Omega_{i0} = \bar{\rho}_{i0}/\rho_{\rm{c0}}$, $\Omega_{\varphi0}= 1-\Omega_{m0} - \Omega_{r0} - \Omega_{\nu0}$, $\kappa\rho_{\rm{c0}} = 3H_0^2$ and $a = 1/(1+z)$ is the scale factor ($z$ is the redshift). As first pointed out in Ref.~\cite{DeFelice:2010pv}, not all of the Galileon parameters (in our case $c_2$, $c_3$ and $\xi$) are independent because of a scaling degeneracy. For instance, on plugging the tracker relation into Eqs.~(\ref{eq:density-background}) it can be noted that the resulting expressions do not change under the transformations $c_2 \rightarrow c_2/B^2$, $c_3 \rightarrow c_3/B^3$ and $\xi \rightarrow \xi B$, for any constant $B$. This holds for all physical quantities, including the perturbed ones. Fixing one of the model parameters is the easiest way to break the scaling degeneracy. The most natural way of doing so is to fix $c_2 = -1$, so that $\mathcal{L}_2$ becomes the standard kinetic energy term, but with the opposite sign. By plugging the tracker relation into the present-day Friedmann equation and the Galileon field equation of motion, we obtain the constraints: $\xi =\sqrt{6\Omega_{\varphi0}}$ and $c_3 = {1}/\left({6\sqrt{6\Omega_{\varphi0}}}\right)$. In this way, {\it the physics of the Galileon model is completely specified by $\Omega_{\varphi0}$ (note $c_2 = -1$), with no free functions to tune such as, for instance, in $f(R)$ gravity} \cite{Sotiriou:2008rp} (see also \cite{Motohashi:2012wc, He:2013qha, Baldi:2013iza}). The Galileon model studied here therefore contains the same free parameters as $\Lambda\rm{CDM}$, although it has different background dynamics and gravitational interactions.

Here, we focus on the large-scale structure constraints that can be derived using linear perturbation theory. In this regime, the modifications to the growth of structure induced by the Galileon field can be captured by defining an effective time-dependent gravitational strength, which is given by:

\bq
&&\frac{G_{\rm eff}}{G}(a)  =\nonumber \\
&&1 + \frac{\Omega_{\varphi0}}{(H(a)/H_0)^4}\left[1 - 2\frac{\dot{H}(a)}{H(a)^2} - \frac{\Omega_{\varphi0}}{(H(a)/H_0)^4}\right]^{-1}.
\eq
For later stages of structure formation, the picture becomes more complex due to the nonlinearities of the screening mechanism \cite{Barreira:2013eea, Barreira:2013xea, Barreira:2014zza}.

\section{Methodology}

Our results were obtained with the publicly available {\tt CAMB} \cite{camb_notes} and {\tt CosmoMC} \cite{Lewis:2002ah} codes, both modified for Galileon cosmologies \cite{Barreira:2012kk, Barreira:2013jma}. In addition to $\Sigma m_\nu$, we fit: the physical energy density of baryonic matter and cold dark matter, $\Omega_{b0}h^2$, $\Omega_{c0}h^2$, respectively; the approximate CMB angular acoustic scale $\theta_{\rm MC}$ (this is a {\tt CosmoMC} parameter); the optical depth to reionization $\tau$; and the scalar spectral index $n_s$ and amplitude $A_s$ (at  $k = 0.05\ \rm{Mpc}^{-1}$) of the primordial power spectrum. We also quote constraints on the {\it rms} linear matter fluctuations at $8\ {\rm Mpc}/h$, $\sigma_8$, which is a derived parameter. 

We consider three data combinations. The first dataset (denoted {\it P}) comprises the Planck data for the temperature anisotropy power spectrum, including the low-$l$, high-$l$ and low-$l$ combined with WMAP9 polarization data \cite{Ade:2013kta, Ade:2013zuv}. This piece of the likelihood also contains nuisance parameters used to model foregrounds, and instrumental and beam calibrations. The {\it PL} dataset adds to {\it P} the data for the lensing potential power spectrum measured by the Planck satellite \cite{Ade:2013tyw}. At the current level of precision of the CMB lensing data, we can ignore nonlinear corrections on the angular scales probed. Finally, the {\it PLB} dataset also includes the BAO measurements from the 6dF \cite{Beutler:2011hx}, SDSS DR7 \cite{Padmanabhan:2012hf} and BOSS DR9 \cite{Anderson:2012sa} galaxy surveys. 

To illustrate the impact of $\Sigma m_\nu$, we consider two Galileon models: one in which the number of massive neutrinos $N^{\nu}_{\rm massive} = 0$ and $\Sigma m_\nu = 0$ (we call this the base Galileon model), and another for which $N^{\nu}_{\rm massive} = 3$ and $\Sigma m_\nu$ is a free parameter (we call this model $\nucubic$). At the precision of the current data, the impact of the neutrino mass splitting is negligible, and hence, one can assume that the three neutrino masses are quasi-degenerate ($m_1 \simeq m_2 \simeq m_3 \simeq m_\nu > 0.1\ {\rm eV}$). Analogously, we also consider a $\nulcdm$ model, which we use for comparison and to establish a reference to assess the goodness of fit of the Galileon models. We always fix the effective number of relativistic neutrinos $N^{\nu}_{\rm eff} = 3.046$. In future work, we plan to relax this to study the effect of extra relativistic degrees of freedom, e.g., sterile neutrinos.

\begin{figure*}
	\centering
	\includegraphics[scale=0.390]{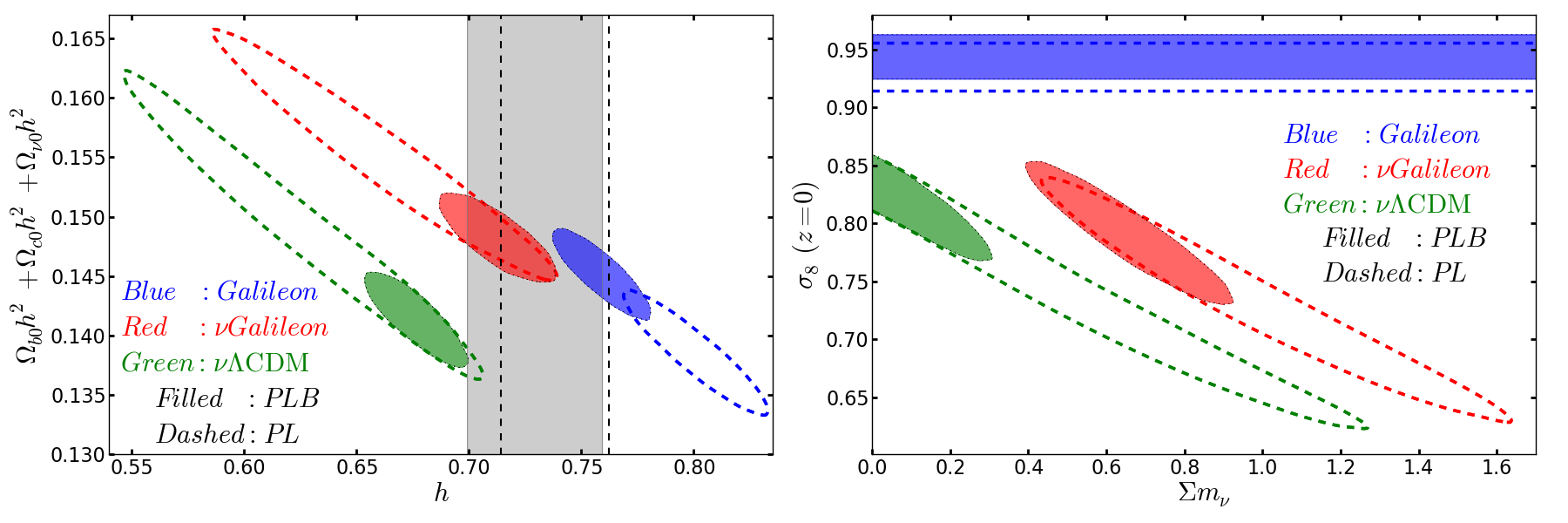}
	\caption{Marginalized two-dimensional $95\%$ confidence limit contours obtained using the {\it PL} (open dashed) and {\it PLB} (filled) datasets for the base Galileon (blue), $\nucubic$ (red) and $\nulcdm$ (green) models. In the left panel, the vertical bands indicate the $68\%$ confidence limits of the direct measurements of $h$ presented in Ref.~\cite{Riess:2011yx} (open dashed) and Ref.~\cite{Humphreys:2013eja} (grey filled). In the right panel, the horizontal bands indicate the $95\%$ confidence interval on $\sigma_8$ obtained using the {\it PL} (open dashed) and {\it PLB} (blue filled) datasets for the base Galileon model (which does not contain massive neutrinos).}
\label{fig:contours}\end{figure*}

\begin{figure}
	\centering
	\includegraphics[scale=0.310]{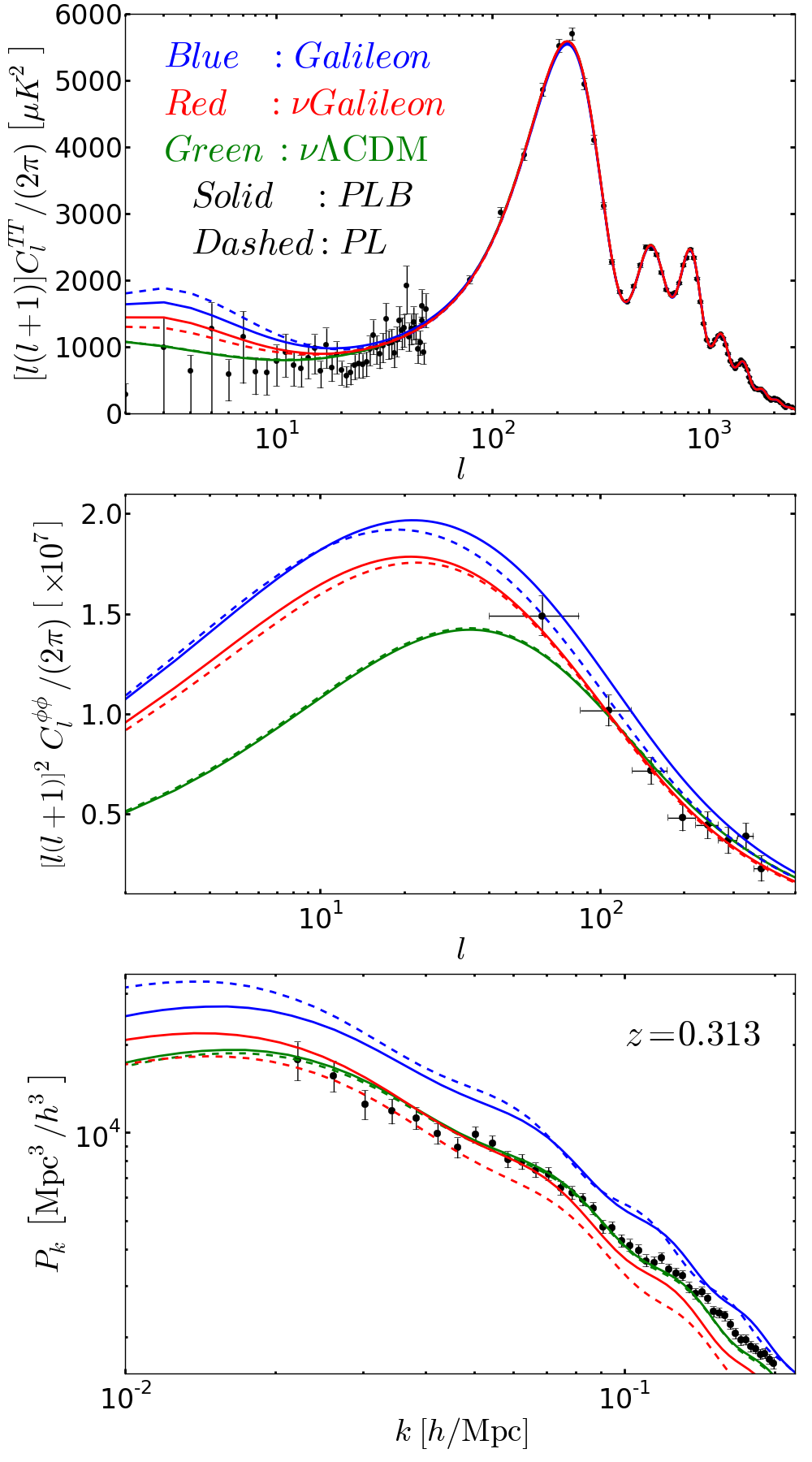}
	\caption{CMB temperature anisotropy (top), CMB lensing (middle) and linear matter power spectra (bottom) of the best-fitting base Galileon (blue), $\nucubic$ (red) and $\nulcdm$ (green) models for the {\it PL} (dashed) and {\it PLB} (solid) datasets.  In the upper and middle panels, the data points show the power spectrum measured by the Planck satellite \cite{Ade:2013zuv, Ade:2013tyw}. In the lower panel, the data points show the SDSS-DR7 Luminous Red Galaxy host halo power spectrum as presented in Ref.~\cite{Reid:2009xm}, but scaled down to match approximately the amplitude of the best-fitting $\nucubic$ ({\it PLB}) model.}
\label{fig:bfs}\end{figure}

\begin{table*}
\caption{Summary of the one-dimensional marginalized likelihood distributions. The upper part of the table shows the best-fitting $\chi^2 = -2\rm{ln}\mathcal{L}$ values (where $\mathcal{L}$ is the likelihood) of the components of the {\it P}, {\it PL} and {\it PLB} datasets. The goodness of fit of the Galileon models can be inferred by comparing the respective $\chi^2$ values with $\nulcdm$, which has been shown to be a good fit to these data in \cite{Ade:2013zuv}. The lower part of the table shows the $1\sigma$ limits on the cosmological parameters obtained for the {\it PL} and {\it PLB} datasets ($h$ and $\sigma_8$ are derived parameters).}
\begin{tabular}{@{}lccccccccccc}
\hline
\hline
\rule{0pt}{3ex}  Parameter/Dataset  & \ \ ${\rm Base\ Galileon}$ & $\nucubic$ &  $\nulcdm$ &\ \ 
\\
\hline
\rule{0pt}{3ex} $(\chi^2_{P} ; --  ; --) $                                   &\ \  $(9829.8\ ; -- \ ; -- )$        &  \ \ $(9811.5\ ; -- \ ; -- )$      &  $(9805.5\ ; -- \ ; -- )$&\ \ 
\\
$(\chi^2_{P} ; \chi^2_{L} ; -- )$                 &\ \  $(9834.6\ ; 8.0\ ; --)$      &  \ \ $(9811.6\ ; 4.4\ ; --)$    &  $(9805.3\ ; 8.8\ ; --)$ &\ \ 
\\
$(\chi^2_{P} ; \chi^2_{L} ; \chi^2_{B})$                              &\ \  $(9834.6\ ; 22.2\ ; 8.0)$ &  \ \ $(9813.5\ ; 4.5\ ; 1.0)$ &  $(9805.4\ ;8.7\ ; 1.4)$ &\ \ 
\\
\hline
\rule{0pt}{3ex}$100\Omega_{b0} h^2$: {\ \ \ ({\it PL, PLB})}                   &\ \  $(2.233 \pm 0.028\ ; 2.177 \pm 0.024)$ &  \ \ $(2.161 \pm 0.030\ ; 2.194 \pm 0.024)$ & $(2.182 \pm 0.035\ ; 2.214 \pm 0.025)$ \ \ 
\\
$\Omega_{c0} h^2$: {\ \ \ \ \ \ \ \ \  ({\it PL, PLB})}          &\ \  $(0.116 \pm 0.002\ ; 0.124 \pm 0.002)$ &  \ \ $(0.123 \pm 0.003\ ; 0.119 \pm 0.002)$ & $(0.121 \pm 0.003\ ; 0.118 \pm 0.002)$ \ \ 
\\
$10^4\theta_{\rm MC}$: {\ \ \ \ \ \ ({\it PL, PLB})}                  &\ \  $(104.17 \pm 0.061\ ; 104.05 \pm 0.058)$ &  \ \ $(104.04 \pm 0.066\ ; 104.10 \pm 0.056)$ & $(104.08 \pm 0.073\ ; 104.14 \pm 0.057)$ \ \ 
\\
$\tau$: {\ \ \ \ \ \ \ \ \ \ \ \ \ \ \ \ \ ({\it PL, PLB})}             &\ \  $(0.067 \pm 0.011\ ; 0.052 \pm 0.010)$ &  \ \ $(0.087 \pm 0.012\ ; 0.088 \pm 0.013)$ & $(0.091 \pm 0.013\ ; 0.092 \pm 0.013)$ \ \ 
\\
$n_s$: {\ \ \ \ \ \ \ \ \ \ \ \ \ \ \ ({\it PL, PLB})}                   &\ \  $(0.970 \pm 0.007\ ; 0.952 \pm 0.006)$ &  \ \ $(0.948 \pm 0.009\ ; 0.960 \pm 0.006)$ & $(0.954 \pm 0.009\ ; 0.963 \pm 0.006)$ \ \ 
\\
${\rm ln}(10^{10}A_s)$: {({\it PL, PLB})}                            &\ \  $(3.034 \pm 0.020\ ; 3.019 \pm 0.019)$ &  \ \ $(3.085 \pm 0.023\ ; 3.081 \pm 0.024)$ & $(3.093 \pm 0.024\ ; 3.090 \pm 0.024)$ \ \ 
\\
$\Sigma m_\nu \ [\rm{eV}]$: {\ \ ({\it PL, PLB})}             &\ \  $(0\ {\rm fixed}            \ ;  0\ {\rm fixed})$                               &  \ \ $(0.980 \pm 0.237\ ; 0.651 \pm 0.106)$ & $( < 0.551\ ;  < 0.127)$ \ \ 
\\
\rule{0pt}{3ex}$h$: {\ \ \ \ \ \ \ \ \ \ \ \ \ \ \ \ \ ({\it PL, PLB})}                  &\ \  $(0.800 \pm 0.013\ ; 0.758 \pm 0.009)$ &  \ \ $(0.663 \pm 0.030\ ; 0.712 \pm 0.010)$ & $(0.634 \pm 0.036\ ; 0.677 \pm 0.009)$ \ \ 
\\
$\sigma_8(z = 0)$: {\ \ ({\it PL, PLB})}                              &\ \ $(0.935 \pm 0.010\ ; 0.944 \pm 0.010)$ &  \ \ $(0.733 \pm 0.042\ ; 0.792 \pm 0.025)$ & $(0.757 \pm 0.056\ ; 0.816 \pm 0.018)$ \ \ 
\\
\hline
\hline
\end{tabular}
\label{table:chi2}
\end{table*} 
\section{Observational constraints} 

Figure \ref{fig:contours} shows marginalized two-dimensional $95\%$ confidence level contours and Table \ref{table:chi2} summarizes the likelihood statistics for different data combinations. The best-fitting CMB temperature spectrum, CMB lensing spectrum and linear matter power spectrum are shown in Figure ~\ref{fig:bfs}. As indicated by the larger values of $\chi^2$, the base Galileon model provides a much poorer fit to the data than $\Lambda\rm{CDM}$ models. This become even clearer as more datasets are considered. In particular, this model fails to provide a reasonable fit to any of the {\it PLB} datasets, e.g., $\chi^2_{\it Lensing} = 22$, for 8 degrees of freedom; and $\chi^2_{\it BAO} = 8$ for 3 degrees of freedom (it is not straightforward to quote the number of degrees of freedom for the CMB temperature data, due to the way in which the low-$l$ data are analysed). 

These observational tensions can be understood as follows. The angular acoustic scale of the CMB fluctuations is given by $\theta^* = r_s^*/d_A^*$, where $r_s^* = \int_{z_*}^\infty {c_s{\rm d}z}/{H(z)}$ and $d_A^* = \int_0^{z_*} {{\rm d}z}/{H(z)}$ are, respectively, the sound horizon and the comoving angular diameter distance at the redshift of recombination $z_*$; $c_s = 1/\sqrt{3\left(1 + 3\rho_b/(4\rho_\gamma)\right)}$ and $\rho_b$ and $\rho_\gamma$ are the energy densities of baryons ($b$) and photons ($\gamma$). The constraints on $\theta^*$ are practically the same in the Galileon and $\Lambda\rm{CDM}$ models, since its value is related to the CMB acoustic peak positions, which makes it essentially model independent. For fixed cosmological parameters, $H(a)$ is the same at early times in the Galileon and $\Lambda\rm{CDM}$ models (cf.~Eq.~(\ref{eq:tracker_H})). Hence, $r_s^*$ is also the same. At late times, however, $H(a)$ is smaller in the Galileon model than in $\Lambda\rm{CDM}$ (cf.~Eq.~(\ref{eq:tracker_H})), due to the phantom nature of the tracker solution, $w_\varphi < -1$ \cite{DeFelice:2010pv}. The lower expansion rate increases $d_A^*$, which lowers $\theta^*$. In order to compensate for this and preserve the peak positions, the CMB temperature data prefer higher values of $h$ for the base Galileon model. Adding the CMB lensing data slightly lowers the matter density to reduce the amplitude of the predicted lensing power spectrum (although not shown, the best-fitting $C_l^{\phi\phi}$ of the base Galileon model for the {\it P} dataset is similar to that for the {\it PLB} dataset in Fig.~\ref{fig:bfs}). This increases both $r_s^*$ and $d_A^*$, but it affects the latter more. By the above reasoning, $h$ is further pushed towards larger values (blue dashed in Fig.~\ref{fig:contours}). This, however, clashes with the preference of the BAO data for lower values of $h$ and higher values of the total matter density (blue filled in Fig.~\ref{fig:contours}). This modifies both $r_s^*$ and $d_A^*$ to maintain the observed acoustic scale, but has also an impact on the amplitude of the CMB temperature and lensing spectra, which triggers shifts in $n_s$,  $A_s$ and $\tau$ to optimize the fit. However, all of these shifts in the parameters do not lead to a perfect compensation, which results in the poorer fit of the base Galileon model. In particular, in addition to the poor fit to the BAO measurements, the best-fitting base Galileon model to the {\it PLB} dataset overpredicts the measured lensing potential power spectrum and has an excess of ISW power in the low-$l$ region of the CMB temperature spectrum (solid blue lines in Fig.~\ref{fig:bfs}). The latter is caused by a rapid late-time deepening of the gravitational potentials induced by the Galileon field.

The presence of massive neutrinos effectively raises the total matter density today, which increases $H(a)$ at late times (cf.~Eq.~(\ref{eq:tracker_H})). A larger value of $\Sigma m_\nu$ can therefore mimic the effects of increasing $h$ on the value of $d_A^*$ (note that $\Omega_\nu \propto \Sigma m_\nu$). In Fig.~\ref{fig:contours}, it is shown that, if $\Sigma m_\nu$ is a free parameter, then the {\it PL} dataset no longer prefers high values for $h$ (red dashed). This eliminates the tension with the BAO data, as indicated by the overlap of the {\it PL} and {\it PLB} contours (red filled) for the $\nucubic$ model (compared with the corresponding mismatch found for the base Galileon model). The goodness of the fit of $\nucubic$ becomes also substantially better for all the data combinations, as indicated by the $\chi^2$ values of Table~\ref{table:chi2}. The modifications introduced by the massive neutrinos cause the gravitational potentials to deepen less rapidly with time, which reduces the ISW power at low-$l$ (red curves in Fig.~\ref{fig:bfs}). Compared to $\nulcdm$, there is still a slight excess of ISW power, but the larger errorbars on these scales do not allow more stringent constraints to be derived. The fit to the lensing power spectrum is also much better in the $\nucubic$ case, relative to the base Galileon model. Compared to $\nulcdm$, $\nucubic$ yields a slightly better fit, as it predicts a higher amplitude for $C_l^{\phi\phi}$ in the range of multipoles $l \sim 40-80$ and the power decreases more rapidly at higher multipoles. It is noteworthy that for $l \lesssim 40$, the $\nucubic$ and $\Lambda\rm{CDM}$ models make very distinct predictions. As a result, future CMB lensing measurements on these larger angular scales have the potential to distinguish between these two scenarios. For completeness, we note that allowing for $\Omega_k < 0$ also lowers $d_A^*$, and as a result, can also ease the tension between the CMB and the BAO data. The impact of non-zero $\Omega_k$ on the ISW effect and CMB lensing is left for future work.

We do not include data from type Ia Supernovae (SNIa). However, as a test, we have checked the impact of adding the data from the three year sample of the Supernova Legacy Survey (SNLS) \cite{Guy:2010bc} to the {\it PLB} dataset. We find that the SNIa data slightly refines the constraints, without shifting the confidence contours from their central values. The relative goodness-of-fit of the $\nucubic$ and $\nulcdm$ models barely changes.

\section{Discussion} We now discuss the impact that additional data can have in further constraining the parameter space of the $\nucubic$ model.

The vertical bands in the left panel of Fig.~\ref{fig:contours} show the $1\sigma$ limits on $h$ obtained using Cepheid variables reported in Refs.~\cite{Riess:2011yx} (open dashed) and \cite{Humphreys:2013eja} (grey filled). We opted not to include these data in our constraints, since the systematic uncertainties on these measurements are not yet completely understood (see e.g. Ref.~\cite{Efstathiou:2013via} for a discussion). Nevertheless, taken at face value, the $\nucubic$ model is consistent with these measurements. Adding a prior for $h$ to the {\it PLB} dataset would then favour $\nucubic$ over $\nulcdm$. 

The release of the Planck data has revealed an additional tension faced by $\Lambda\rm{CDM}$ models concerning the normalization of the matter density fluctuations. Specifically, the value of $\sigma_8$ inferred from the CMB temperature and lensing data seems to be larger than the values inferred from galaxy lensing or cluster number counts \cite{Ade:2013zuv, Ade:2013lmv}. Recently, Refs.~\cite{Wyman:2013lza, Battye:2013xqa} have shown that the inclusion of sufficiently massive neutrinos can improve the fit of $\Lambda\rm{CDM}$ to these data, but some residual tension remains. Figure ~\ref{fig:contours} shows that the values of $\sigma_8$ in the $\nucubic$ model are substantially smaller than in the base Galileon model, being comparable to those of $\nulcdm$ (see also the bottom panel of Fig.~\ref{fig:bfs}). It is therefore of interest to check whether or not the above tension in $\Lambda\rm{CDM}$ is also present in $\nucubic$. This requires a proper modelling of nonlinear structure formation (e.g. modelling of small scale clustering and halo mass function) in $\nucubic$ cosmologies, which is left for future work (see e.g. \cite{Barreira:2013eea, Barreira:2014zza, Hellwing:2014nma} for steps in this direction).

The high energy part of the Tritium $\beta$-decay spectrum offers a robust way to directly measure neutrino masses in a model independent way. The MAINZ and TROITSK experiments have determined $\Sigma m_\nu \lesssim 6.6\ {\rm eV}$ (at $2\sigma$), but upcoming experiments such as KATRIN will be able to improve the sensitivity to $\Sigma m_\nu \lesssim 0.6\ \rm{eV}$ (see e.g. \cite{Drexlin:2013lha} for a recent review). If light neutrinos are Majorana particles and provide the dominant contribution, neutrinoless double $\beta$-decay experiments will be able to achieve even higher precision, probing completely the quasi-degenerate spectrum for which $\Sigma m_\nu \gtrsim 0.3\ \rm{eV}$ (see e.g. \cite{Vergados:2012xy} for a recent review). All these experiments are expected to reach their forecast sensitivity in a few years time. This will bring the terrestrial constraints on $\Sigma m_\nu$ into a regime where they can be used to further test the $\nucubic$ model, for which $\Sigma m_\nu \gtrsim 0.4\ {\rm eV}$ (at $2\sigma$) using cosmological data (c.f~Fig.~\ref{fig:contours}).

The $\nucubic$ model predicts a negative sign for the ISW effect due to the late time deepening of the gravitational potentials. This result is at odds with the current observational suggestions that the sign of the ISW effect is positive (see e.g.~\cite{Giannantonio:2012aa, Ade:2013dsi, Ferraro:2014msa}), as it is in $\Lambda\rm{CDM}$. However, some skepticism has been raised about some of these observational results \cite{HernandezMonteagudo:2012ms, Cai:2013ik, Francis:2009pt, Francis:2009ps, HernandezMonteagudo:2009fb, Sawangwit:2009gd, LopezCorredoira:2010rr}. Furthermore, in the $\nucubic$ model there is also the additional role that the Vainshtein mechanism may play in alleviating a potential observational tension. This requires more complete modelling of nonlinear structure formation, and as such, it is left for future work.

\ 

In conclusion, the $\nucubic$ model emerges as a simple and attractive alternative to $\Lambda\rm{CDM}$ that is testable with future cosmological and particle physics experiments. Further studies of this model will be of interest not only in understanding better the role that massive neutrinos can play in modified gravity theories, but also in the planning and interpretation of the results from ongoing and future observational missions.

\

\underline{{\it Acknowledgments}} We thank Marco Baldi and David Weinberg for useful comments and suggestions. We are grateful to Antony Lewis for help with the {\tt CosmoMC} code, and to Lydia Heck for valuable numerical support. This work used the DiRAC-2 Data Centric system at Durham University, operated by the Institute for Computational Cosmology on behalf of the STFC DiRAC HPC Facility (\url{www.dirac.ac.uk}). This equipment was funded by BIS National E-infrastructure capital grant ST/K00042X/1, STFC capital grant ST/H008519/1, and STFC DiRAC Operations grant ST/K003267/1 and Durham University. DiRAC is part of the National E-Infrastructure. AB is supported by FCT-Portugal through grant SFRH/BD/75791/2011. BL is supported by the Royal Astronomical Society and Durham University. This work has been partially supported by the European Union FP7  ITN INVISIBLES (Marie Curie Actions, PITN- GA-2011- 289442) and STFC (ST/F001166/1).

\bibliography{cubic-massivenu.bib}

\end{document}